\documentclass{kluwer}    
\usepackage{psfig}
\usepackage[]{graphics}

\newdisplay{guess}{Conjecture}

\begin{document}                                                                                   
\begin{article}
\begin{opening}         
\title{The First Detection of Cobalt in a Damped Lyman Alpha System.} 
\author{Sara L. \surname{Ellison}}  
\institute{European Southern Observatory, Chile}
\author{Sean G. \surname{Ryan}}  
\institute{Open University, UK}
\author{Jason X. \surname{Prochaska}}  
\institute{Carnegie Observatories, USA}
\runningauthor{Sara L. Ellison et al.}
\runningtitle{Cobalt in DLAs.}
\date{April 12, 2001}

\begin{abstract}
We present the first ever detection of Cobalt in a Damped Lyman Alpha
system (DLA) at $z \sim 2$.  In addition to providing important clues to
the star formation history of these high redshift galaxies, we discuss
how studying the Co abundance in DLAs may also help to constrain models
of stellar nucleosynthesis in a regime not probed by Galactic stars.
\end{abstract}

\end{opening}           

\section{Introduction}  

The study of elemental abundances in Damped Lyman Alpha systems 
(DLAs) at high redshift represents one of our best opportunities 
to probe galaxy formation and chemical evolution at early times.  
By coupling measurements made in high $z$ DLAs with our knowledge of 
abundances determined locally and with nucleosynthetic models,
we can start to piece together the star formation histories of 
these galaxies.  To this end, certain key elements are generally targeted,
for example Zn and Cr are traditionally used as a measure of the 
overall metallicity and gas-to-dust ratio in DLAs, whereas Si and 
S are used to determine
whether there is any evidence for $\alpha$ element overabundances,
indicative of SN Type II enrichment.  Here, we propose that Co,
which has never before been featured in DLA studies, yet is well-studied
in Galactic stars, may be an additional element that can provide useful
clues to the nature of star formation at high redshift.

\section{Why Study Cobalt?}

The intriguing trends of [Co/Fe]
as a function of [Fe/H] in the various Galactic stellar
populations indicate that cobalt may be an element
that can provide important clues to chemical evolution, see Figure 1.  
In sub-solar metallicity thin disk stars, Co is found to be slightly
underabundant with respect to Fe ([Co/Fe] $\sim -0.1$, 
Gratton and Sneden 1991).   This trend
continues in moderately metal-poor halo stars down to metallicities of
[Fe/H] $\sim -2.5$.  For more metal-poor halo stars, a considerable
overabundance of Co with respect to Fe is observed, $0 <$ [Co/Fe] $< 0.8$,
(McWilliam et al. 1995; Ryan, Norris and Beers 1996).  Interestingly, this
coincides with a downturn in relative Cr to Fe abundances, such that Co and
Cr are closely anti-correlated in the metal poor population. 
Similar [Co/Fe] overabundances are observed in the bulge 
([Co/Fe] $\sim +0.3$, McWilliam \& Rich 1994), which, we recall, 
is as old as the halo, and to a lesser degree in the
thick disk ([Co/Fe] $\sim +0.1$, Prochaska et al. 2000).

The detection of Co~II transitions is a challenging possibility for
ISM absorption line spectroscopy, both locally and at high $z$.  
Although there are numerous transitions
that lie at convenient rest wavelengths for both local and high
redshift studies, only two interstellar measurements currently exist
(e.g. Mullman et al 1998).
One of the advantages of moving to higher redshifts is that the relatively
strong UV lines become shifted into the optical where one has
access to more efficient detectors and larger telescopes.  In addition,
the observed equivalent width is
increased by a factor of ($z + 1$).  However, the fundamental problem  
remains that the Co~II lines are all intrinsically weak and 
that the solar abundance of Co is $8 \times 10^{-8}$ that
of hydrogen.

\begin{figure}
\centerline{\rotatebox{0}{\resizebox{10cm}{!}
{\includegraphics*[20,350][585,770]{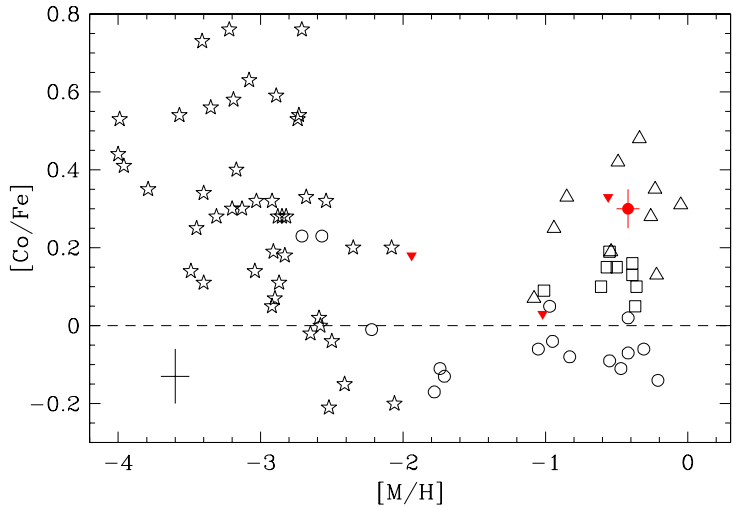}}}}
\caption{Relative cobalt abundances as a function of metallicity for 
different Galactic populations and DLAs: 
open stars --- metal-poor halo stars from McWilliam et al (1995) and Ryan, 
Norris, \& Beers (1996);
open circles --- moderately metal-poor thin disk and halo stars from Sneden \& 
Gratton (1991); open triangles --- bulge stars from McWilliam \& Rich (1994);
open squares --- thick-disk stars from Prochaska et al. (2000).
The solid circle represents Q2206-199 and the down-turned filled triangles are 
upper limits for three other DLAs (Q1223+17 [this work], HE1104$-$1805 [Lopez
et al 1999] and Q302$-$223 [Pettini et al 2000]).  
The cross in the lower left corner
of the figure shows the typical error bar for stellar points.}
\end{figure}

\section{Cobalt in DLAs}

In order to maximise the likelihood of detecting Co, we have selected
two DLAs with high $N$(Fe) that lie in front of relatively bright
background QSOs.  The former criterion is to enable us to more easily
reach our target detection limit of [Co/Fe] $\le$ 0, which will be sufficient
to distinguish between most of the populations plotted in
Figure 1.  Selecting bright QSOs obviously allows us to reach the requisite
S/N in feasible amounts of time.  The two systems chosen for this pilot
study are Q2206$-$199 ($B = 17.8, z_{em} = 2.559$, $z_{abs} = 1.92$) 
and Q1223+17 ($B = 18.5, z_{em} = 2.936$, $z_{abs} = 2.466$).  Observations
were obtained with UVES (on the VLT) and HIRES (on Keck) the details of
which (along with data reduction steps and complete abundance analysis)
are described in full by Ellison, Ryan and Prochaska (2001).  Here, we
limit ourselves to reporting the final determined Co abundances which
are [Co/H] = $-0.51$ for Q2206$-$199 and [Co/H] $< -1.76$ for Q1223+17.
These points are plotted in Figure 1 along with Galactic stellar
values and additional DLA limits derived from the literature.

\section{Results}

We have shown that the detection of Co in DLAs is feasible with
echelle spectrographs on 8-m class telescopes.
The first detection of Co in a DLA appears to be consistent with the
abundances measured in the Galactic bulge and although this doesn't
necessarily equate DLAs with high redshift bulges, it may indicate
that the two share some similarities in their star formation
histories.  However, the interpretation of these results is hindered
by our lack of understanding of the nucleosynthetic processes that
produce Co.  Only recently have models been formulated that can successfully
reproduce Galactic [Co/Fe] trends by forcing a deeper mass cut or higher
explosion energies (Nakamura et al 1999, 2000).  
However, matching the [Co/Fe] has come at the expense 
of high Ni abundances which are not seen in stars.  In the case of
Q2206$-$199, however, we $do$ find that high [Co/Fe] is accompanied by
high [Ni/Fe].  Therefore, although the results here represent a
somewhat preliminary study of Co with a clear need for more data points,
we have demonstrated the interest of studying this element both in
the context of understanding better the nature of DLAs and for
providing a regime for the testing of nucleosynthetic models.

\acknowledgements
The authors would like to express their gratitude to their colleagues 
Jacqueline Bergeron, Patrick Petitjean and Max Pettini for their consent
to use some of the data presented here.

\end{article}
\end{document}